\documentclass[preprint,showpacs,preprintnumbers,amsmath,amssymb]{revtex4}

\usepackage{graphicx}
\usepackage{dcolumn}
\usepackage{bm}

\newcommand{\half}{\frac{1}{2} }

\newcommand{\ket}[1]{|#1 \rangle}
\newcommand{\braket}[2]{\langle #1 |#2 \rangle}
\newcommand{\ketbra}[2]{| #1 \rangle \langle #2 |}

\newcommand{\complex}{\mathbf{C}}

\newcommand{\Hilbert}{\mathcal{H} }
\newcommand{\Kilbert}{\mathcal{K} }

\newcommand{\Tr}{\mathrm{Tr}}

\newcommand{\vphi}{\varphi}

\newcommand{\migi}{\rightarrow}

\newcommand{\rem}[2]{\noindent\textbf{Remark #1.#2.}\\}

\newcommand{\etal}{\textit{et al.}}

\begin{document}

\preprint{APS/123-QED}

\title{New derivation of Born's law and parameter estimation\\ based on a relative state formulation}

\author{Fuyuhiko Tanaka}
\email{ftanaka@stat.t.u-tokyo.ac.jp}
 \altaffiliation[Also at ]{Department of Mathematical Informatics , University of Tokyo.}
 \affiliation{%
PRESTO, Japan Science and Technology Agency, 4-1-8, Honcho Kawaguchi, Saitama, Japan
}%

\date{\today}

\begin{abstract}
Excluding the concept of probability in quantum mechanics,
we derive Born's law from the remaining postulates in quantum mechanics using type method.  
We also give a way of determining the unknown parameter in a state vector based on an indirect measurement model. 
While Deutsch adopt a concept of rational decision-maker who introduces probability,
 we adopt a concept of statistician in our measurement model and clarifies the distinguished feature of quantum measurement.
Like many worlds interpretation, our scenario gives a simple solution for problem of measurement.
\end{abstract}

\pacs{03.65.-w,03.65.Ca,03.65.Ta}

\maketitle

\section{\label{sec:level1}Introduction}

Classical mechanics is ultimately deterministic and concept of 
probability is secondary thing and regarded as a useful tool.
In quantum mechanics, if we do not consider any measurement process,
then a state vector in a Hilbert space of a closed system evolves in a deterministic way with time subject to a unitary operator.

In traditional Copenhagen interpretation, measurement process is described in the following way~\cite{vN1955}.
Physical measurement device is written as a projection-valued measurement (PVM). 
The probability of obtaining a measurement result is calculated by PVM and a state vector.
 But any deterministic prediction is impossible in principle
 except for the eignstates of an observable.
After we obtain the measurement result, the state jumps to the corresponding state vector.
However, it should be possible to describe the whole process including measurement and macroscopic observation in a quantum mechanical setting
because ideally, both the macroscopic measurement device and observer are collection of microscopic subsystems, each is described in quantum mechanics.
Here, we call such theory \textit{pure quantum mechanics (pure QM)}. 
Lots of people have investigated whether the measurement process is really described in pure QM formulation described by using only state vector and unitary operation.


Among lots of approaches, Everett~\cite{Everett1957} realized a wave function or state vector is \textit{relative}.
According to his idea, 
 after a measurement, the world branches into ones with each measurement result 
and these worlds do not affect each other, the whole process is unitary. 
This idea is called many worlds interpretation (MWI), but his original idea seems to be misunderstood and
thus unacceptable to many people.
In the present paper, we mainly agree with the original idea of relative state formulation by Everett 
and try to explain a measurement process as a unitary process in a more convincing way with some modification.

Important claims by Everett are as follows:
\begin{enumerate}
\item Measurement process is only a unitary process of a closed system and measurement postulate is unnecessary theoretically. 
\item Collapse of the wave function never happens anyway, 
but is justified by considering the cascade of the measurement process and relative state.
\item Wave function or state vector does not have an absolute meaning but 
a clue for an experimental prediction by an observer. Thus, the form of a wave function depends on where the observer is.
\item Probability is a secondary concept even in quantum mechanics.
\item No difference exists between an object to be measured and a measurement device.
\item No difference exists between macroscopic and microscopic system.
\end{enumerate}

In the present paper, following the above claims, we derive a discrete version of Born's law from other non-probabilistic postulates in pure QM.
Derivation of Born's law was given by several authors decades ago~\cite{DeWittGraham1973}.
Among such works, Deutsch~\cite{Deutsch1999} focused on a game theoretic viewpoint.
His main idea is to adopt a concept of \textit{rational decision maker} who introduces probability through the evaluation of the value of a certain game,
 which is a way of introducing probability in classical statistics. 
Our idea is to adopt a concept of \textit{statistician}, which is similar to Deutsch, 
but inference by statistician may depend on another criteria like symmetry argument~\cite{Everett1957} or his own belief. 
Statistician does not distinguish a formal interference of two orthogonal states and a classical probabilistic mixture of two states
 under a single type of measurement.
Thus, statistician also considers the hypothesis testing of a pure state after the estimation of the unknown coefficients,
  which is the most distinguished feature in pure QM.
  In principle, we can design a measurement device which detects a superposition state in the perfect efficiency.
In addition, our motivation is not to investigate quantum cosmology, 
but to describe a measurement process as a unitary process in detail and clarify theoretical limitation concerned with 
quantum estimation like determining the unknown coefficients in a state vector.

Among previous works, our philosophy is very similar to Hartle~\cite{Hartle1968}.
But there are at least three different points.
First, we adopt an indirect measurement model, which is more suitable to discuss a state change after measurement.
Second, we emphasize reversibility of the whole measurement process.
Third, we use a modern information theoretical tool called \textit{type method}, which brings us the most elaborate result among previous works. 
Not only do we propose the new interesting concepts, but also give stimulating topics in various fields of physics 
 such as general relativity, quantum field theory, quantum information, quantum statistics, and mathematical physics.


Structure is as follows.
In the next section, first we review postulates of pure QM and Born's law in the present paper.
Then, we consider the whole measurement process as a unitary process, which is essentially the same as Everett.
Next, Born's law is derived using modern technique of information theory.
We also compare our method with Everett's original result and describe testing model. 
In the next section, we discuss some foundational problems such as 
Schr\"{o}dinger's cat, the collapse of the wave function and problem of measurement.

%



\section{A way of determining the unknown coefficients based on type method}

Suppose that the unknown wave function $\psi (x) $ is expanded by
 an orthonormal system of known functions $\{\vphi_{i} \}$:
\[
 \psi(x) = \sum_{i=1}^{k}c_{i} \vphi_{i}(x),
\]
where $k$ is an arbitrary large number and $c_{1},\dots, c_{k} \in \complex $.
When we do not have any probabilistic law or postulate in quantum mechanics, how does a statistician determine
 the unknown parameter $c_{i}$?
It seems impossible on first sight, but it is shown to be possible if we admit performing an experiment infinitely many times. 
For simplicity, we consider two level system or qubit system $\Hilbert = \complex^2$.
Then as usual in quantum information, an orthonormal system is given by $\{ \ket{0}, \ket{1} \}$.
We expand the unknown state vector of unit length $\ket{\psi} \in \Hilbert$ as
\[
 \ket{\psi} = c_0 \ket{0} + c_1 \ket{1}
\]
using the two parameter $c_{0}, c_{1} \in \complex$.
We give one way of determining the absolute value $|c_{0}|, |c_{1}|$ using only pure quantum mechanical (QM) postulates.
Although QM postulates are originated from von Neumann~\cite{vN1955}, 
here let us cite another set in a more simple and modern form by Nielsen and Chuang~\cite{Nielsen}. \\

\noindent \textbf{Postulate 1.}\\ 
Associated to any isolated physical system is a complex vector space with inner product (that is, a Hilbert space) known as
 the \textit{state space} of the system. The system is completely described by its \textit{state vector}, which is a 
 unit vector in the system's state space.\\

\noindent \textbf{Postulate 2.}\\ 
The evolution of a \textit{closed} quantum system is described by a \textit{unitary transformation}.
That is, the state $\ket{\psi}$ of the system at time $t_1$ is related to the state $\ket{\psi'}$ of the system at time
$t_2$ by a unitary operator $U$ which depends only on the times $t_1$ and $t_2$, 
\[
 \ket{\psi'} = U\ket{\psi}. 
\]

\noindent \textbf{Postulate 3.}\\
The state space of a composite physical system is the tensor product of the state spaces of the component physical systems.
Moreover, if we have systems numbered $1$ through $n$ , and system number $i$ is prepared in the state $\ket{\psi_i}$, 
then the joint state of the total system is $\ket{\psi_1}\otimes \ket{\psi_2} \otimes \cdots \ket{\psi_{n}}.$ \\

As we mentioned before, we excluded Postulate of Measurement.
In the above setting, Born's law claims that the probability of obtaining $j \in \{0, 1\}$ with the state $\ket{\psi}$ is given by
\begin{equation}
Prob (j | \psi ) =  | \braket{j}{\psi} |^{2} \label{eq:Born}
\end{equation}
when the measurement of the observable $X = 0 \ketbra{0}{0} + 1 \ketbra{1}{1}=\ketbra{1}{1}$ is performed.
One of our objective is to \textit{derive} the above relation~(\ref{eq:Born}) from the rest of Postulates.
We also indicate that 
probability in the above formula is a formal representation of uncertainty to an observer.

\subsection{Measurement process}


In pure QM, the whole system is represented by a state vector and
 we do not admit any concept of a probabilistic mixture or statistical ensemble.  
Indeed, a density matrix can be purified due to purification and a nonunitary process can be
 realized as a unitary process in a larger Hilbert space~(See, e.g., Chap. 2 and Chap. 8 in Nielsen~and~Chuang~\cite{Nielsen}).
Mainly following Everett~\cite{Everett1957}, we describe a measurement process as a unitary operation.
Since his model of the measurement process is too simple, we introduce an indirect measurement model below.

Denote a microscopic system to be measured as $S$, 
the environmental system as $E$, and the system of a measurement device (apparatus) system
 as $A$.
 New concept is a \textit{statistician} $F$~(named after a famous statistician R.~A.~Fisher).
 He performs an experiment and guess the unknown parameter.
 He also estimates and predicts a result with the additional assumption of probability.
In our framework, probability is introduced only by him. 
Nature in quantum mechanics is also deterministic as Everett pointed out.


Our measurement model composes two step.
First, $S$ and $A$ are coupled and information of $S$ 
is copied to $A$.
This may be a thermal irreversible process and described by trace preserving and completely positive map.
But, in our ideal case, this process is also represented as a unitary (thus, reversible) process
 by getting together with the environmental system $E$.

Second, $F$ is coupled with $A$ and $F$ reads out registered information of $A$.
Practically, each state vector of the register system $A$ is distinguishable in a macroscopic level
 like scintillator or photon detector etc.
This process is also unitary (and reversible !!) when we consider the very huge environmental system $E$. 
The setup that $F$ is not directly coupled with $S$ 
admits various possibilities of describing phenomena such as post-measurement state of $S$, measurement error of $A$, etc.

For reader's convenience we distinguish between the system to be measured and the system of a measurement device 
but they could be swapped.
In our definition, essentially, there is no distinction between a macroscopic system and a microscopic system,
 or an object to be measured and a device to measure,
although we can distinguish whether the measurement system is good for experimenter or not.
We agree with Everett in this point.

Usually, a macroscopic system is a very large one, which would not be written practically.
However, in the following argument, it is enough to consider a certain subsystem.
For simplicity of calculation, we adopt two-level system as $S$ and $A$.
The environmental system and a system of statistician himself are also implicitly 
assumed to be described in finite-level systems.

Now we explain two step measurement process in detail.
In order to grasp an explicit image, suppose that a unitary process begins with $t=0$ and ends at $t=t_{1}$.
A composite system $\Hilbert_{SAE}$ is closed and $F$ does not interact with them during the first step of the measurement process.
Due to Postulate 2, the time evolution of the whole system $\Hilbert_{SAE}$ is written as
\begin{eqnarray*}
\ket{\Psi_{in}}_{SAE}  &:= & \ket{\psi(t=0)}=(c_{0}\ket{0}_{S} + c_{1}\ket{1}_{S}) \otimes \ket{0}_{A} \otimes \ket{0}_{E} \\
\ket{\Psi_{out}}_{SAE} &:= & \ket{\psi(t=t_{1})} =
 U(t_{1}) \ket{\psi(t=0)} = c_{0}\ket{0}_{S} \otimes \ket{0}_{A} \otimes \ket{f_0}_{E}+ c_{1}\ket{1}_{S} \otimes \ket{1}_{A} \otimes \ket{f_1}_{E}
\end{eqnarray*}

Hereafter, we often omit $\otimes $.
We also omit $t_1$ because it is a fixed constant in the following argument.
The above unitary process may be relevant with the environmental system $E$, 
like thermal energy exchange or other inessential process.
For simplicity, we assume that there are no cross terms $ \ket{0}_{S}\ket{1}_{A}$, $\ket{1}_{S}\ket{0}_{A}  $
 because it would bring a measurement error but not essential in the following argument.
If the final states of $E$ vary according to the state vector of the register system $A$, 
it implies in principle the environmental system $E$ also holds (partial) information of the state of $S$.
 In this case, dividing $E=E_1 +E_2$, we replace $A$ with $A+E_{1}$ and take $E_{2}$ as the environmental system.
 Anyway, if we choose a measurement device $A$ in a good manner, we can assume that $E$ has no information on the states of 
 $S$, in other words, the final state of the environmental system is the same state $\ket{f}_{E}$.
Thus, the final state is given by
\[
  \ket{\Psi_{out}}_{SAE} = \left\{ c_{0}\ket{0}_{S}\ket{0}_{A} + c_{1}\ket{1}_{S}\ket{1}_{A} \right\} \otimes \ket{f}_{E}. 
\]
From now on, we omit the environmental system.

Now the first part of the measurement process is complete, and $F$ reads out the measurement result
 using a unitary evolution between an apparatus system $A$ and himself. 
(We neglect the environmental system.)
It is trivial to describe the unitary evolution; (just substitute $S \migi A$, $A \migi F$.)
Final state is as follows:
\[
  \ket{\Psi_{out}}_{SAF} =  c_{0}\ket{0}_{S}\ket{0}_{A}\ket{0}_{F}  + c_{1}\ket{1}_{S}\ket{1}_{A}\ket{1}_{F}. 
\]
Some readers may be concerned about the chain of measurement as Wigner pointed out~\cite{Wigner1963}.
However, as we shall see later, the chain ends to $F$.

In the conventional way, the above process is explained as follows.
We have a microscopic system $S$ represented as $ \ket{\psi} = c_{0} \ket{0} + c_{1} \ket{1}$.
Then, statistician $F$ performs the measurement of an observable $X$,
where $X \ket{0} =0 $ and $X \ket{1} = \ket{1}$.
Probability of obtaining $j$ is evaluated as $|c_{j}|^{2}$ \textit{due to Postulate of Measurement}.
Practically this description is very convenient.
However, its simplicity prevents us from investigating how the physical object is represented 
or whether the difference between an object and a measurement device exists or not.
Note that a generalized measurement or a positive operator valued measurement~(POVM) is also described by an observable or a PVM in a larger Hilbert space due to Naimark extention (See, e.g., Peres~\cite{Peres1995}).

\subsection{Mathematical model of stable measurement device}

In the above section, we assume that the usual macroscopic measurement process is described as a unitary one.
In pure QM, we do not distinguish between microscopic systems and macroscopic systems
and we do not admit any probabilistic model even in the environmental system.
Readers may wonder if a measurement process is really reversible.
Thus, before proceeding our main result, we show that a deterministic and reversible unitary evolution could be seen as a nonunitary irreversible process.
Such a unitary process is said to be \textit{almost irreversible}.

The following is a mathematical model of a stable measurement device and an example of almost irreversible process.
Consider a single atom which has two energy levels, i.e., the ground state and the excited state. 
First, an atom in the excited state $\ket{e}_{S}$ emits a single photon in the environmental system $\ket{0}_{E} $ and jumps to the ground state $\ket{g}_{S}$:
\[
 \ket{e}_{S}\ket{0}_{E} \longrightarrow  \ket{g}_{S}\ket{\hbar \nu}_{E}. 
\]
Then, the photon in the environmental system is absorbed in the $L$ site spin system $A$, and induces $L/2$ spin flips:
\[
 \ket{\hbar\nu}_{E} \ket{\uparrow \uparrow \dots \uparrow }_{A} \longrightarrow \ket{0}_{E} \ket{\uparrow \downarrow \dots \uparrow \downarrow }_{A}. 
\]
Spin exchange happens sequentially as
\begin{eqnarray*}
 \ket{0}_{E} \ket{\uparrow \downarrow \uparrow \downarrow \dots \downarrow }_{A} &\longrightarrow & \ket{0}_{E} \ket{\downarrow\uparrow \uparrow \downarrow \dots \downarrow }_{A} \\
  &\longrightarrow & \ket{0}_{E} \ket{\downarrow\uparrow \downarrow \uparrow  \dots \downarrow }_{A} \\
  &\longrightarrow & \dots, 
\end{eqnarray*}
each of which can be written as a unitary process.
This process is deterministic but indistinguishable in a macroscopic level.
After $\binom{L}{L/2} $ flips, again they emit the single photon with energy $\hbar \nu$,
\[
  \ket{0}_{E} \ket{\downarrow\uparrow \downarrow \uparrow  \dots \uparrow }_{A} 
 \longrightarrow \ket{\hbar\nu}_{E} \ket{\uparrow \uparrow \uparrow  \dots \uparrow }_{A}. 
\]
The above states are at least distinguished by the measurement of total spin, $S_{tot} := \sum_{j}S_{j}$.
That is, the large number of states is regarded as $\ket{1}_{A}$ like
\begin{eqnarray*}
\ket{\uparrow  \dots \uparrow \uparrow }_{A} & \cdots & \ket{0}_{A} \\
\ket{\downarrow\uparrow \downarrow \uparrow  \dots \uparrow }_{A} & \cdots  & \ket{1}_{A} \\ 
 & \vdots &  \\
\ket{\downarrow\uparrow \downarrow \uparrow  \dots \downarrow }_{A}& \cdots & \ket{1}_{A}
\end{eqnarray*}
This kind of measurement device is subject to a unitary process as a whole and reversible.
As the size of spin site goes large, the number of spin flip $\binom{L}{L/2}$ increases exponentially and
 the state $\ket{1}_{A}$ seems very stable even in our macroscopic time scale when each spin flip happens during very short time $\Delta t$.
Moreover, such a process may be coupled with another similar process in turn.
Thus, the whole process, which is unitary and reversible, is regarded as an irreversible process to us.
Time arrow in this way could arise and irreversible measurement processes could happen in a macroscopic way.

In the present paper, thus, probabilistic phenomena are removed in the ultimate level.
Only when statistician estimates something, a concept of probability is introduced as a tool.
The other processes are all described in a deterministic way. 
This is a spirit of Einstein; he said \textit{Der Alte wurfelt nicht}.

\subsection{Repetition of measurement process}

Now, we explain how the statistician $F$ reads results and estimates the unknown parameter in the above setup.
Statistician $F$ prepares individual $N$ systems, which are identified with each other.
We assume that the initial state vectors and unitary operations are in the same form
 and there is no interaction between one and another system.


Thus, due to Postulate of composite system (Postulate $3$), 
the final state of the measurement process is written as
\begin{equation}
 \ket{\Psi_{out}^{(N)}}:= U^{\otimes N} \ket{\Psi_{in}}^{\otimes N}
 = \ket{\Psi_{out}}^{\otimes N}, \label{eq:product}
\end{equation}
where 
\[
\ket{\Psi_{out}}= c_0 \ket{0}_{S}\ket{0}_{A} + c_1 \ket{1}_{S}\ket{1}_{A}. 
\]
For simplicity, choosing the phase of $\ket{0}_{A}$ and $\ket{1}_{A}$, we take $c_{0}$ and $c_{1}$ as positive constants.
Normalization implies $c_0^2 + c_1^2=1$.
Note that these two constants are coefficients of a state vector concerning the orthonormal basis $\{ \ket{0}_{S}, \ket{1}_{S} \} $.
At most, they present the extent of the interference of two vectors.


We do not require all measurement processes start simultaneously.
If we recover the initial condition of the experimental setup, we can measure only one system at one time in repetition.
In addition, mathematically, we do not care about the physical realization of a qubit.
For example, we could use photon qubits in the half of total repetition and nuclei qubits in the other half.

Now we proceed detailed calculation.
When $N=2$, the right hand side of (\ref{eq:product}) is expanded in the following way. From now on, the indices of the system are often omitted
 and $\ket{i}_{S}\ket{j}_{A}$ is often written as $\ket{ij}$ shortly.
\begin{eqnarray*}
(c_0 \ket{00} + c_1 \ket{11})^{\otimes 2} 
&=&  c_0^2  \ket{00} \ket{00}  +  c_1^2 \ket{11}\ket{11}
  + c_0 c_1 \ket{00}\ket{11} + c_1 c_0 \ket{11} \ket{00}  \\
 &=& c_0^2  \ket{00}^{\otimes 2} +c_1^2  \ket{11}^{\otimes 2} 
 + \sqrt{2}c_0 c_1 S( \ket{00}\ket{11} ),  
\end{eqnarray*}
where $S( \ket{00}\ket{11} ) $ is a symmetrized and normalized vector.
In general form, a symmetrized and normalized vector is defined by
\[
 S(\ket{00}^{\otimes m}\ket{11}^{\otimes N-m} ) :=
  \binom{N}{m}^{-\half} \{ \ket{00}^{\otimes m}\ket{11}^{\otimes N-m}+ (\text{permutation terms}) \}.
\]
Setting $p$ and $q$ as $p=c_0^2$, $q=1-p=c_1^2$, we obtain 
\begin{eqnarray*}
 \ket{\Psi_{out}}^{\otimes N}&=& \sum_{m=0}^{N}
  p^{\frac{m}{2}} q^{\frac{N-m}{2}} \binom{N}{m}^{\half} S(\ket{00}^{\otimes m}\ket{11}^{\otimes N-m} ).
\end{eqnarray*}


We see that the $N$ tensor product state of the final state is expanded using $N+1$ orthonormal symmetrized state vectors $S(\ket{00}^{\otimes m}\ket{11}^{\otimes N-m} )$,
 where each coefficient is given by
\[
p^{\frac{m}{2}} q^{\frac{N-m}{2} } \binom{N}{m}^{\half} .
\]
Maximum coefficient is given by $m_{*}= [ Np ]$, where $[\cdot ]$ denotes Gauss's symbol and hereafter often omitted.
Now let us consider the asymptotic behavior with $N \migi \infty$. 
Using the \textit{type method}, which is familiar with classical information theory~\cite{Cover2005},
 we can show that the corresponding term becomes dominant and the other coefficients of the symmetrized 
state vectors become \textit{exponentially} small.
This is a stronger result than previous works~\cite{Everett1957, Graham1973, Hartle1968}. 
Thus, we obtain in the asymptotic setting, 
\begin{eqnarray*}
 \ket{\Psi_{out}}^{\otimes N}&=& \sum_{m=0}^{N} p^{\frac{m}{2}} q^{\frac{N-m}{2}}  \binom{N}{m}^{\half}     S(\ket{00}^{\otimes m}\ket{11}^{\otimes N-m} )\\
& \approx  & S(\ket{00}^{\otimes m_{*}}\ket{11}^{\otimes N-m_{*}} ) \\
&=& S(\ket{00}^{\otimes Np}\ket{11}^{\otimes Nq} ). 
\end{eqnarray*}
The above formula is essential in our argument.
This kind of idea is also used in quantum information~\cite{Schumacher1995}.

\subsection{Derivation of Born's law}

First, suppose that statistician $F$ knows the coefficients $c_0$ and $c_1$.
Then, without any observation of measuring device $A$,
 $F$ easily expects that the number of $\ket{0}$ is near $Np$ and the number of $\ket{1}$ is near $Nq$.
When $N \migi \infty$, his guess holds true.
Without the concept of probability, just two expansion coefficients of a state vector
 are given an operational meaning.
Statistician cannot predict each outcome but can predict relative frequency of measurement outcome.
If the relative frequency of $j$ is interpreted as a probability of obtaining an outcome $j$ in a single measurement
 for the prepared system represented by $\ket{\psi}$,
 its probability is given by
 \[
  Prob (j| \psi) = |c_{j}|^{2}.
 \]
Thus, Born's law~(\ref{eq:Born}) is derived in pure QM.
Although we consider only a two-valued measurement in the above argument, 
we can easily extend to multivalued-measurement in the same line.

Now we return to the original problem about the wave function. 
 If we take Dirac's delta functions $\{ \ket{x} \}$ as a basis, 
and formally expand a wave function as
\[
 \ket{\psi} \approx \sum_{x} \ket{x} \braket{x}{\psi} =\sum_{x} c(x) \ket{x} ,  
\]
then we would obtain the operational meaning of the coefficients $c(x)$.
This leads to usual Born's law, which claims that 
the probability of finding a particle around $x$ is proportional to the square of a wave function $ |c(x)|^{2}$. \\


\rem{2}{1}
Outside the whole experimental setup, we do not know which individual device obtains the measurement outcome $0$ or $1$.
Thus, symmetrized vector state appears in a formal way.
In classical independently identical distributed (i.i.d.) trials of coin toss, 
exchanging the order of each toss does not affect the probabilistic model.
Our symmetrized vector state is the pure quantum mechanical analogue.\\

\rem{2}{2}
Of course, if $F$ knows $c_0=0, c_1=1$ in advance,
even when $N=1$, without any observation of experimental setup,
 $F$ predicts $\ket{\Psi_{out}}= \ket{1}_{S}\ket{1}_{A} $ perfectly.
 


\subsection{Reading measurement outcome and guess}

Next, suppose that statistician $F$ read out the measurement result.
It does not matter whether he knows two coefficients or not.
In this situation, $F$ interacts the tensor product system of the register system $A$.
Thus, we write a state vector of $F$ himself from the outside viewpoint denoted as $G$,
 which never interacts the whole system $\Hilbert_{SAF}$.


Denoting the measurement results as $i^{N} := (i_1,\dots, i_{N})$, where $i_{1},\dots, i_{N} \in \{0, 1\} $,
$2^{N}$ orthonormal vectors $\ket{i^{N}}_{F}$ are defined.
They represent the states of $F$ just after reading-out the measurement result $i^{N}$.
The initial state is denoted as $\ket{\Omega }_{F}$ and orthogonal to the other state vectors.
Now the whole state vector representing the state just before reading out the measurement result $i^{N}$ is written as
\[
\ket{\Psi_{in}}:=   S(\ket{00}_{SA}^{\otimes Np}\ket{11}_{SA}^{\otimes Nq} ) \otimes \ket{\Omega }_{F}.
\]
Since $S$ and $A$ is entangled, we cannot omit the $S$ part.
We assume that there is no classical or quantum error in reading out process.
Then, the whole state vector representing the state just after reading out the measurement result $i^{N}$
 is written as
\begin{eqnarray*}
\ket{\Psi_{out}} = U_{AF}\ket{\Psi_{in}}
&=&S(\ket{00}_{SA}^{\otimes Np}\ket{11}_{SA}^{\otimes Nq}\ket{00\cdots 011\cdots 1}_{F}) \\
&=& S(
 (\ket{0}^{\otimes Np}\ket{1}^{\otimes Nq})_{S} 
 (\ket{0}^{\otimes Np}\ket{1}^{\otimes Nq})_{A}
 (\ket{0}^{\otimes Np}\ket{1}^{\otimes Nq})_{F}
 ). 
\end{eqnarray*}
$F$ reads out the information of the registered system $A$ through the unitary process.
After the unitary process, $F$ is in the state of perceiving, say, $0010\cdots 01$.
Finally, statistician $F$ guesses using the above data and other deduction in the following way.
As $N$ becomes large enough, each coefficient of a state vector becomes extremely small in the order of $2^{-N/2}$.
State vectors registering $Np$ $0$s and $Nq$ $1$s are dominant.
The Hilbert space representing the whole system $SAF$ is almost equal to the subspace spanned by $ \{ \ket{i^{N}}_{S}\ket{i^{N}}_{A} \ket{i^{N}}_{F} \} $,
 where $i^{N}$ is a sequence composing of $Np$ $0$s and $Nq$ $1$s.
Thus, $F$ can expect that he is in the only one classical world governed by classical probability theory
 and $F$ observes one \textit{typical sequence} composing of $Np$ $0$s and $Nq$ $1$s by chance.
 (In the above experiment, $F$ cannot predict the order of $0$s and $1$s.)
In terms of MWI, infinitely many divided worlds are almost collected again and composes one world. (See, Graham~\cite{Graham1973}.) 
At least, when $F$ believes that he is in a typical world, 
he counts the number of $0$s and $1$s and can estimate $c_{0}$ and $c_{1}$ as below:
\[
 \hat{c}_0 = \sqrt{ \frac{m_{0}}{N}},\quad \hat{c}_{1} = \sqrt{1 - \frac{m_{0}}{N}}, 
\]
where $m_{0}$ denotes the count of $0$s, 
and $\hat{c}_{j}$ denotes an estimate of the unknown parameter $c_{j}$.
When $N \migi \infty $, the estimates are true.
It would be possible for $F$ to estimate using other prior information, his own belief, strategy, or criteria as a statistician
 when $N$ is not large enough.\\

\rem{2}{3}
Due to time invariance, we do not need simultaneous experiments.
We do not need to perform even real experiments.
When considering infinitely many imaginary experiments plus one real experiment,
 it brings the concept of ``statistical ensemble".
Statistician $F$ expects that the measurement outcome $0$ will be registered with relative frequency $p=|c_{0}|^2$
 \textit{before reading the measurement result}.
Now using this information, 
$F$ can estimate the probability of obtaining the registered outcome in the only one real experiment 
\text{before reading the measurement result}.
This is the justification of statistical ensemble based on pure QM.\\

\subsection{Testing model}

After the estimation of the unknown parameter, $F$ has to test his estimate like the following way.
He prepares a new unitary process $U_{SA}'$ between $S$ and $A$ such that
\[
\left\{ 
\begin{matrix}
\ket{\psi}_{S}\ket{0}_{A}  & \longrightarrow & \ket{\psi}_{S}\ket{0}_{A} \\
\ket{\psi^{\perp}}_{S}\ket{0}_{A}  & \longrightarrow & \ket{\psi^{\perp}}_{S}\ket{1}_{A} \\
\end{matrix}
\right. ,
\]
where $\ket{\psi^{\perp}} $ denotes an orthogonal state to $\ket{\psi}$, say,
\[
\ket{\psi^{\perp}} := c_{1} \ket{0} - c_{0} \ket{1}.
\]
The above unitary process is possible when both $\{ \ket{0}_{S}, \ket{1}_{S} \}$ and $\{ \ket{0}_{A}, \ket{1}_{A} \}$ belong to the same irreducible representation of $\mathrm{SO}(3)$.
We leave the reading process unchanged.
Then, after performing the new measurement, it is expected that $F$ only sees the $0$ in register system $A$.
If he obtains $1$, then he must check the measurement device, object, his guess, misspecified model, and assumption of each unitary process etc.
Even when he does not obtain any outcome except for $0$, he cannot be confident since $N$ is finite.
Again, a certain probabilistic model is introduced in order to assess his hypothesis in a quantitative way.
This testing procedure is overlooked in previous works~\cite{Everett1957, Hartle1968, DeWittGraham1973, Deutsch1999}.

Here, we mention a difference between conventional QM and pure QM. 
In usual formulation (based on Copenhagen interpretation), the whole estimation process is described in two conceptually different manners.
One claims that we have a quantum state written as a density matrix $\rho = c_{0}^{2} \ketbra{0}{0} + c_{1}^{2} \ketbra{1}{1}$
and perform a projective measurement $\{ \ketbra{0}{0}, \ketbra{1}{1} \}$.
The other claims that we have a quantum state written as a density matrix 
$\rho = \ketbra{\psi}{\psi} =  c_{0}^{2} \ketbra{0}{0} + c_{1}^{2} \ketbra{1}{1} 
+ c_{0}c_{1}\ketbra{0}{1} +   c_{1}c_{0}\ketbra{1}{0}$ and perform the same measurement.
Then, they have to argue which is really correct when they consider macroscopic interference like the Schr\"{o}dinger's cat.

In pure QM, a probabilistic mixture of a physical reality is not allowed.
The both density matrices are recognized as formal representations of the uncertainty to the statistician $F$.
In a single fixed measurement, we do not distinguish both representations.
On the other hand, 
pure QM is originally based on the nonrelativistic Schr\"{o}dinger equation and is a very flexible mathematical tool, 
therefore allows nonphysical phenomena to be described.
Thus, we see whether the above testing process is possible in principle or not.
If impossible in principle, then the superposition of orthogonal states $\ket{0}_{S}$ and $\ket{1}_{S}$ is regarded as a formal one.


\subsection{Many worlds interpretation}

In the next section, we show that $F$ \textit{could} consider the collapse of the wave function happens 
and the initial state vector of the system $S$ jumps to the state vector $\ket{0010\cdots 01}_{S}$.
However, from $G$ outside the composite system $SAF$, such different states interfere with each other
 and this reading out process never causes the collapse of wave function as a whole. 
This idea is originally introduced in order to avoid the collapse of a wave function and to keep unitarity of the whole measurement process in pure QM 
by Hugh Everett~\cite{Everett1957} (Many Worlds Interpretation).
Everett considered such a measurement (both measurement process and reading out process in our setting) brings a branching process.
According to him, the world from the viewpoint of $F$ inside the whole system $SAF$ branches into $2^{N}$ different worlds.

Unfortunately, among some people, this original idea seems to be misunderstood as creating new worlds after local measurement processes.
We emphasize that the world from the standpoint of the external observer $G$ never divides.
There are much more than $2^{N}$ orthonormal state vectors in $\Hilbert_{SAF}$, which are \textit{neglected during the measurement process}.  
Only one state vector of the world is 
spanned by such huge number of vectors and is unitarily rotated.
The whole evolution from the viewpoint of $G$ is reversible, although returning back to the initial state is impossible in an ergodic sense.

Another comment is on a practical difference between macroscopic system and microscopic system.
In principle, both of them are not distinguished.
As a consequence, it is inevitable to describe a superposition in a macroscopic system in order to describe a measurement process
 in pure QM setting.
In the above setting, the system including $F$ are written as a superposition of orthonormal vectors like
\[
 \ket{\Psi_{out}} = \frac{1}{\sqrt{2^{N}}} \sum_{i_{N}\in \{0,1\}^{N}} \ket{i_{N}}_{SAF}
\]
from the external observer $G$.
However, for fixed $N$ and a macroscopic system like $F$, 
this can be interpreted as a formal representation of the uncertainty to $G$, a macroscopic observer,
 because pure QM does not admit a concept of probability (at least in our framework).
For example, a throw of a classical dice is represented as
\[
 \ket{\Psi} = \frac{1}{\sqrt{6}} (\ket{1} +\ket{2}  +\ket{3}  +\ket{4}  +\ket{5}  +\ket{6}),
\]
where $\ket{j}$ denotes that $F$ sees $j$ after a throw of the dice.
In infinitely many trials, as discussed above, we obtain a typical sequence where each relative frequency is equal to $1/6$.
In a macroscopic level, we cannot recognize directly the interference between the original orthogonal states. 
This kind of restriction is imposed on a unitary process.
Thus, in a classical approximation, a stochastic process is also represented by a branching world.
For example, three runs of dice throw, we need to prepare at least the following orthonormal vectors:
\[
 \ket{\Omega \Omega \Omega}, \ket{i\Omega\Omega}, \ket{ij\Omega}, \ket{ijk}, \quad 1 \leq i,j,k \leq 6,
\]
where $\Omega$ denotes the state before $F$ sees the number.
Then, usual sequential process is represented as follows.
First, we set an initial state:
\[
 \ket{\Psi} = \ket{\Omega\Omega\Omega}.
\]
After the first throw of the dice, we obtain
\[
 \ket{\Psi} = \frac{1}{\sqrt{6}} \sum_{i=1}^{6}\ket{i\Omega\Omega}.
\]
Then, sequentially we obtain
\begin{eqnarray*}
 \ket{\Psi} &=& \frac{1}{\sqrt{6^2}} \sum_{i=1}^{6}\sum_{j=1}^{6}\ket{ij\Omega}, \\
 \ket{\Psi} &=& \frac{1}{\sqrt{6^3}} \sum_{i=1}^{6}\sum_{j=1}^{6}\sum_{k=1}^{6}\ket{ijk}.
\end{eqnarray*}
In this sense, it is possible to say that our world branches into $6^{3}$ different worlds
 after three throws.
Empirically, such a process is recognized as a probabilistic model by usual macroscopic people
 because we do not have the ability of detecting the superposition.



\subsection{Comparison with Everett's method}

In our framework of quantum mechanics, we exclude any concept of probability assuming that the whole setting is ideal.
This restriction is the same as deterministic classical mechanics.
As Everett pointed out, we can argue in a more quantitative way by adopting a certain probabilistic model.
In the above setting, Everett derived a probability measure from the additivity requirement and 
its probability measure brings usual Born's law.
From Bayesian viewpoint, this deduction corresponds to the determination of an objective prior distribution of the unknown parameter
 and the choice of an objective prior is arbitrary even in classical Bayesian statistics (See, e.g., Robert~\cite{Robert}).

Mathematically speaking, his idea to derive Born's law is also insufficient.
He introduced the above probability measure in the branching process.
In his setting, an observer performs measurement and reads outcome \textit{sequentially},
 and estimates the expectation of an observable.
However, it is known that this kind of sequential definition causes a contradictory problem.
In two-valued measurement, suppose that we observe $0$ or $1$ sequentially in repetition of the measurement of one system.
We can perform two lines of repetition, 
$A_1 := (a_1,a_2, a_3, \dots, ),\quad A_2 := (b_1, b_2, b_3, \dots )$, where $a_j, b_j \in \{0,1\}$.
Then, we take product of each sequence $c_i := a_i b_i$.
If we have a certain pair of elaborate sequences (See, e.g., Williams~\cite{Williams}), then both average
$ \sum_{i}\frac{a_{i}}{N}, \sum_{i}\frac{b_{i}}{N}$ converge and
$ \sum_{i}\frac{c_{i}}{N}$ never converges. 
Everett avoids this technical problem by introducing the above probability measure.
On the other hand, our method is based on properties of the tensor product,
that is, we used the type-method in classical information theory. 
If we take $N \migi \infty$, infinitely many $0$s and $1$s appear,
 but the proportion of both counts is finite and determined, 
thus, statistician $F$ can define this proportion as a probability and obtain Born's law.
Our way is both conceptually and technically different from Everett's one. 
We are not certain that our method is extended to infinite-dimensional cases in a straightforward way,
 and such topics are attractive and new in the field of mathematical physics.\\



\section{Cascaded measurement process}


In the above framework, considering counting numbers argument in the cascaded measurement process,
 we explain the discrete version of the collapse of a wave function.

First, we define a relative state vector as an analogue of conditional distribution in classical probability theory.
This concept is also attributed to Everett~\cite{Everett1957}.
On the tensor product of two Hilbert spaces, $\Hilbert \otimes \Kilbert$,
let a state vector written as
\[
 \ket{\Psi} = \sum_{\alpha}u_{\alpha} \otimes e_{\alpha},
\]
where $u_{\alpha} \in \Hilbert$ are unnormarized vectors and $\{ e_{\alpha} \} \subset \Kilbert $ is an orthonormal system.
Then, a \textit{relative state vector} with respect to $e_{\alpha} $ is a vector in $\Hilbert $ defined as 
\[
\left.  \ket{\Psi}\right|_{\alpha} = \frac{u_{\alpha}}{ || u_{\alpha} || }.
\]

\subsection{Schr\"{o}dinger's cat}

Schr\"{o}dinger's cat is one  of the most  famous thought experiments in quantum physics.
A microscopic superposition is connected with a macroscopic system in this sort of experiment.
For simplicity, we only write the state of so-called half-alive and half-dead of a cat,
\[
\ket{\psi} = \frac{1}{\sqrt{2}} \{ \ket{\textrm{L}}_{C} + \ket{\textrm{D}}_{C} \},
\]
where L denotes the cat is alive and D denotes the cat is dead. 
When sticking to physical reality, this kind of state seems to be difficult to understand.
Thus, lots of ideas and explanations have been proposed.
One of them is the decoherence caused by the environmental system (for example, see Nielsen and Chuang, Chap.8~\cite{Nielsen}.). 
Very roughly speaking, they take another state vector, and write the state vector of the cat and an environmental system as
\[
\ket{\psi} =
 \frac{1}{\sqrt{2}} 
\{ \ket{\textrm{L}}_{C}\otimes \ket{0}_{E}  
 +
   \ket{\textrm{D}}_{C} \otimes \ket{1}_{E}  
\}.
\]
Then, they claim that the cat state is in the mixed state  
\[
 \rho_{C} := \Tr_{E} [ \ketbra{\psi}{\psi} ] = \half \ketbra{\textrm{L}}{\textrm{L}} + \half \ketbra{\textrm{D}}{\textrm{D}}.
\]
However, a system of one real cat and the ensemble of imaginary infinite samples are confused in the above picture.
The cat alive may claim that ``I am alive, their estimated state is wrong and the correct state is $\ket{\psi} = \ket{\textrm{L}}$."
Not surprisingly, the cat alive also knows the microscopic state is not in the ground state. 
This is like an interacting free detection by Kwiat \etal~\cite{Kwiat1995}, 
and the following measurement process like self-reference could happen inside in our framework.
First, we take an additional memory in the cat, which is referring to itself and default is $\ket{\Omega}_{C'}$.
\[
\ket{\Psi_{in}} = \frac{1}{\sqrt{2}} 
     \{ \ket{\textrm{L}}_{C}+ \ket{\textrm{D}}_{C}\} \otimes \ket{\Omega}_{C'},
\]
Then, after referring to itself through a unitary process between $C$ and $C'$, we obtain
\[
\ket{\Psi_{out}} = \frac{1}{\sqrt{2}} 
 \{ \ket{\textrm{L}}_{C}\otimes \ket{\text{I am alive.}}_{C'}  
 +
   \ket{\textrm{D}}_{C} \otimes \ket{\Omega}_{C'}.  
\}
\] 
For this result, a relative state vector with respect to ``detecting alive" is given by
\[
 \left. \ket{\Psi_{out}}\right|_{\text{alive}} =  \ket{\textrm{L}}_{C}.
\]
Thus, Everett's original idea of a relative state vector seems to give a reasonable explanation.
Based on the above insights, we consider the collapse of a wave function in our framework.
Note that we do not bring epistemology and do not require an awakening intelligent cat, rather
 we give an intuitive meaning of the relativity of a state vector.

\subsection{Cascaded measurement}


Now we consider a sequential measurement process $\mathbf{M}_{1}$ and $\mathbf{M}_{2}$ for the system to be measured.
Since the environmental system $E$ is taken common to both processes, we omit it.
First an apparatus system $A$ is coupled with $S$ from $t=0$ to $t=t_1$, then another apparatus system $B$ is coupled with $S$ from $t=t_1$ to $t= t_2$, where $0 < t_1 < t_2$.
After the first measurement process $\mathbf{M}_{1}$, the inside state is:
\[
 \ket{\Psi(t_1)} = (U_{SA}\otimes I_{B}) \ket{\Psi_{in}} = 
(c_{0} \ket{0}_{S}\ket{0}_{A}
 + c_{1} \ket{1}_{S} \ket{1}_{A} ) \ket{0}_{B}
\]
After the second measurement process $\mathbf{M}_{2}$, the inside state is:
\begin{eqnarray*}
\ket{\Psi(t_2)} &=& (U_{SB}\otimes I_{A}) \ket{\Psi(t_1)} \\
 &=& 
c_{0}  ( c'_{0}\ket{0}_{S} \ket{0}_{B} + c'_{1}\ket{1}_{S} \ket{1}_{B} )  \ket{0}_{A}
+
c_{1}  ( c''_{0}\ket{0}_{S} \ket{0}_{B} + c''_{1}\ket{1}_{S} \ket{1}_{B} )  \ket{1}_{A} \\
&=&
\left\{  c_{0} c'_{0} \ket{0}_{S}  \ket{0}_{A} \ket{0}_{B} 
   + c_{0} c'_{1} \ket{1}_{S} \ket{0}_{A} \ket{1}_{B} \right. \\
&&  {} \left. +  
 c_{1} c''_{0} \ket{0}_{S}  \ket{1}_{A} \ket{0}_{B} 
   + c_{1} c''_{1} \ket{1}_{S} \ket{1}_{A} \ket{1}_{B}
\right\}
\end{eqnarray*}
Note that $ |c_{0}|^2 + |c_{1}|^2=1, |c'_{0}|^2 + |c'_{1}|^2=1, |c''_{0}|^2 + |c''_{1}|^2=1$.
Finally, after $N$ runs of experiments, statistician $F$ reads both $A$ and $B$ 
in the product state $\ket{\Psi(t_2)}^{\otimes N}$.
We denote the number of the registered state $\ket{i}_{A} \ket{j}_{B} $ as $m_{ij}$. 
Then, $F$ can estimate the absolute value of the unknown coefficients using the following formula.
$|c_{0} c'_{0}|^2 = m_{00}/N$, $|c_{0} c'_{1}|^2 = m_{01}/N$, $|c_{1} c''_{0}|^2 = m_{10}/N$, and $|c_{1} c''_{1}|^2 = m_{11}/N $.
After reading process, statistician $F$ estimates that 
\begin{eqnarray*}
|c_{0}|^2 = \frac{m_{00} + m_{01}  }{N},& & |c_{1}|^2 = \frac{m_{10} + m_{11}  }{N}, \\
|c'_{0}|^2 = \frac{m_{00}}{m_{00}+m_{01}}, && |c'_{1}|^2 = \frac{m_{01}}{m_{00}+m_{01}}, \\
|c''_{0}|^2 = \frac{m_{10}}{m_{10}+m_{11}}, && |c''_{1}|^2 = \frac{m_{11}}{m_{10}+m_{11}}. 
\end{eqnarray*}

Next, we consider an additional reading out process between the first measurement process and the second one.
Then, we focus on the system including $\ket{0}_{A}$.
\begin{eqnarray*}
 \left( U(t_2 -t_1) \ket{0}_{S} \ket{0}_{A} \ket{0}_{B} \right) ^{\otimes M_{0}}
&=&  \left( c'_{0} \ket{0}_{S} \ket{0}_{B} +  c'_{1} \ket{1}_{S} \ket{1}_{B}  \right) ^{\otimes M_{0}} \ket{0}_{A}^{\otimes M_{0}},
\end{eqnarray*}
where $M_{0}$ denotes the number of $0$s registered in $A$.
We denote the number of counts of $0$ and $1$ as $m_{0|0}$ and $m_{1|0}$.
When $M_{0}$ is large enough,$m_{0|0} \approx M_{0}|c'_{0}|^2$, $m_{1|0} \approx M_{0}|c'_{1}|^2$ holds.
On the other hand, $m_{00} \approx N |c_{0}|^2 |c'_{0}|^2$, $m_{01} \approx N |c_{0}|^2 |c'_{1}|^2$ holds.
In particular, if we take the limit $N, M_{0} \migi \infty $ satisfying $N = M_{0} |c_{0}|^{2}$, then
$m_{0|0} \approx m_{00}$ and $m_{1|0} \approx m_{01}$
holds.
This implies that the above two experiments are equivalent with each other.
In other words, both experiments are not distinguishable in principle.

After the first reading out process of statistician $F$, he composes one of the whole system $\Hilbert_{SAF}$.
Then, from the viewpoint of $F$, any experiment for $\Hilbert_{S}$ is equivalent to the process subject to the unitary evolution $U(t_2 -t_1)$ 
with the initial state vector as a relative state vector with respect to $\ket{j}_{A}\ket{j}_{F}$, that is $\left. \ket{\Psi(t_1) }\right|_{j} =\ket{j}_{S}$.
This argument is the justification of Copenhagen interpretation \textit{only for the statistician $F$ (and $A$)} inside the whole system
 after the reading out process, that is, the interaction with a measurement device.
Collapse of the wave function to $F$ is entirely described in a unitary process 
and the above explanation gives a simple solution for problem of measurement. \\

\rem{3}{1}
$F$ can also perform an adaptive measurement,
that is, the latter measurement process may depend on a measurement outcome registered in $A$.
Then, the latter measurement process is described by $U_{SABE}$.\\



\rem{3}{2}
Some people may be concerned with the above relative state vector.
For example, Jauch \etal~\cite{Jauch1967} says, \\
\\
\textit{
the state vector is reduced to the status of a mere mathematical tool expressing
the part of earlier observations which have relevance for predicting results of later ones.
It thus no  longer represents the $<< $ state $>> $ of an individual system but
 describes only some properties of ensembles of such systems prepared under identical relevant conditions.}\\
\\
However, their comment is misleading.
In classical mechanics, in principle, the system of an object and the system of the observer or measurement device are separated. 
No interaction between them is assumed.
In pure QM, a unitary interaction between 
the system of the object and the system of a measurement device 
is inevitable to obtain an information from one system and we ourselves are not an exception.
Thus, the system of the object could be described in a different way after we obtain information.
Note that the above concept of information is also relative one.
Change of the initial state is recognized as information.

\section{Summary}

We give an operational meaning of the expansion coefficients $c_{i}$ of the state vector concerning an orthonormal basis by using an asymptotic technique for 
the tensor product of $N$ composite systems, which represent a measurement process as a unitary one. 
That is, when one system is represented by
\[
 \ket{\psi}_{S} = \sum_{i=1}^{k} c_{i} \ket{u_{i}},
\]
the unknown $c_{i}$ is estimated using the count of the registered $i$.
This estimate is true only in the limit of $N \migi \infty$.
When $N$ is not large enough, statistician has to infer using his own criterion and prior information.
As in classical statistical decision theory, this inference also depends on the purpose of estimation.
On the other hand, if we know the coefficients $c_{i}$ in advance, 
we can predict the count of the registered $i$
before we see the measurement result.
This estimate is also true in the limit of $N \migi \infty$.
Final step is to test the above estimate or hypothesis, which is usually overlooked.
We prepare the whole system in order that the measurement device $A$ detect the initial state $\ket{\psi}_{S}$ perfectly and never detect the other orthogonal state.
Thus, the estimate seems true unless we see contradictory results.
In this step, for finite $N$, statistician evaluates how confident the result is by using a reasonable probabilistic model.

Theoretically, in a different way from Everett, we derived Born's law 
by the above argument from the postulates of pure QM.
In pure QM, all process including measurement is described as a unitary process from the external observer.
Again, we cite Einstein's famous phrase:\\
\textit{Der Alte wurfelt nicht.}\\


\begin{acknowledgments}
The author was supported by Kakenhi.
The author is also grateful to T.~Sasaki for fruitful discussions and careful reading of the manuscript.
\end{acknowledgments}



\end{document}